\documentclass[aps,pra,twocolumn,showpacs,floatfix,superscriptaddress,nofootinbib]{revtex4-1}

\usepackage{graphicx}%
\usepackage{dcolumn}%
\usepackage{bm}%
\usepackage[colorlinks=true]{hyperref}
\usepackage{color}
\usepackage{amsmath}
\usepackage{amssymb}
\usepackage{braket}
\usepackage[normalem]{ulem}
\usepackage{multirow}
\usepackage{booktabs}
\usepackage{blkarray}
\usepackage{orcidlink}

\newcommand{\changes}[1]{#1}

\renewcommand{\eqref}[1]{Eq.~(\ref{#1})}
\newcommand{\Schr}{Schr\"odinger }
\newcommand{\ip}{I_\mathrm{p}}
\newcommand{\up}{U_\mathrm{p}}
\newcommand{\pb}{\mathbf{p}}

\newcommand{\rb}{\mathbf{r}}

\newcommand{\Ab}{\mathbf{A}}
\newcommand{\Eb}{\mathbf{E}}

\newcommand{\Qprop}{Q{\sc{prop}}}

\begin{document}

\title{Femtosecond pulse parameter estimation from photoelectron momenta using machine learning}

\author{Tomasz Szołdra \orcidlink{0000-0002-2897-0506}}
\affiliation{Doctoral School of Exact and Natural Sciences, Jagiellonian University, \L{}ojasiewicza 11, PL-30-348 Krak\'ow, Poland}
\affiliation{Instytut Fizyki Teoretycznej, Wydzia\l{} Fizyki, Astronomii i Informatyki Stosowanej, Uniwersytet Jagiello\'nski, \L{}ojasiewicza 11, PL-30-348 Krak\'ow, Poland}

\author{Marcelo F. Ciappina \orcidlink{0000-0002-1123-6460}}
\affiliation{Department of Physics, Guangdong Technion - Israel Institute of Technology, 241 Daxue Road, Shantou, Guangdong, China, 515063}
\affiliation{Technion - Israel Institute of Technology, Haifa, 32000, Israel}
\affiliation{Guangdong Provincial Key Laboratory of Materials and Technologies for Energy Conversion, Guangdong Technion - Israel Institute of Technology, 241 Daxue Road, Shantou, Guangdong, China, 515063}
\author{Nicholas Werby \orcidlink{0000-0002-1789-3170}}
\affiliation{Stanford PULSE Institute, SLAC National Accelerator Laboratory 2575 Sand Hill Road, Menlo Park, CA 94025, USA}
\affiliation{Department of Physics, Stanford University, Stanford, CA 94305, USA}
\author{Philip H. Bucksbaum~\orcidlink{0000-0003-1258-5571}}
\affiliation{Stanford PULSE Institute, SLAC National Accelerator Laboratory 2575 Sand Hill Road, Menlo Park, CA 94025, USA}
\affiliation{Department of Physics, Stanford University, Stanford, CA 94305, USA}
\affiliation{Department of Applied Physics, Stanford University, Stanford, CA 94305, USA}
\author{Maciej Lewenstein \orcidlink{0000-0002-0210-7800}}
\affiliation{ICFO-Institut de Ciencies Fotoniques, The Barcelona Institute of Science and Technology, Av. Carl Friedrich Gauss 3, 08860 Castelldefels (Barcelona), Spain}
\affiliation{ICREA, Pg. Lluís Companys 23, 08010 Barcelona, Spain}
\author{Jakub Zakrzewski \orcidlink{0000-0003-0998-9460}}
\affiliation{Instytut Fizyki Teoretycznej, Wydzia\l{} Fizyki, Astronomii i Informatyki Stosowanej, Uniwersytet Jagiello\'nski, \L{}ojasiewicza 11, PL-30-348 Krak\'ow, Poland}
\affiliation{Mark Kac Complex Systems Research Center, Jagiellonian University, \L{}ojasiewicza 11, PL-30-348 Krak\'ow, Poland}
\author{Andrew S. Maxwell \orcidlink{0000-0002-6503-4661}}
\email{andrew.maxwell@phys.au.dk}
\affiliation{Department of Physics and Astronomy, Aarhus University, DK-8000 Aarhus C, Denmark}
\date{\today}

\begin{abstract}
Deep learning models have provided huge interpretation power for image-like data. Specifically, convolutional neural networks (CNNs) have demonstrated incredible acuity for tasks such as feature extraction or parameter estimation. Here we test CNNs on strong-field ionization photoelectron spectra, training on theoretical data sets to `invert' experimental data. Pulse characterization is used as a `testing ground', specifically we retrieve the laser intensity, where `traditional' measurements typically lead to 20\% uncertainty.  We report on crucial data augmentation techniques required to successfully train on theoretical data and return consistent results from experiments, including accounting for detector saturation. The same procedure can be repeated to apply CNNs in a range of scenarios for strong-field ionization. Using a predictive uncertainty estimation, reliable laser intensity uncertainties of a few percent can be extracted, which are consistently lower than those given by traditional techniques. Using interpretability methods can reveal parts of the distribution that are most sensitive to laser intensity, which can be directly associated with holographic interferences. The CNNs employed provide an accurate and convenient ways to extract parameters, and represent a novel interpretational tool for strong-field ionization spectra.
\end{abstract}

\maketitle

\section{Introduction}
Machine learning (ML) has been transformative for science over the last two decades, providing a huge range of new analytical tools. This has affected nearly every avenue of research, with major use across the physical sciences, particularly in particle physics \cite{karagiorgi_machine_2022}, astrophysics \cite{vanderplas_introduction_2012}, and condensed matter physics \cite{carleo_machine_2019, dawid_modern_2022}. Convolutional neural networks (CNNs) have enabled major leaps in computer vision and language processing \cite{rawat_deep_2017, li_survey_2022}. This makes CNNs well-suited for pattern recognition and parameter estimation in scientific data.
For example, CNNs have been used for determining crystal symmetries in electron diffraction \cite{kaufmann_crystal_2020}, and estimating parameters related to gravitational lensing \cite{hezaveh_fast_2017}. However, in the field of strong field physics and attoscience the high interpretability power of CNNs has not been fully explored.

Strong-field physics and attoscience exploit recent advances for producing intense and short laser pulses to image and control matter over attosecond ($10^{-18}$s) timescales \cite{krausz_attosecond_2009, salieres_study_1999, lewenstein_principles_2008, ciappina_attosecond_2017}. These capabilities have led to a wide range of atomic and molecular imaging procedures, e.g., high-order harmonic spectroscopy \cite{itatani_tomographic_2004}, laser-induced electron diffraction \cite{zuo_laserinduced_1996}, photoelectron holography \cite{huismans_timeresolved_2011,figueirademorissonfaria_it_2020}, attosecond streaking \cite{hentschel_attosecond_2001,itatani_attosecond_2002}, and reconstruction of attosecond harmonic beating by interference of two-photon transitions \cite{paul_observation_2001,muller_reconstruction_2002}. However, despite ever more accuracy from experiment and theory, due to the nonlinear nature of the interactions, interpretation of the data is often very challenging. This provides an opportunity for ML methods to be used to extract parameters and physical trends from experimental data sets.

A growing number of studies have begun to use ML techniques for strong-field physics. For example, studies using neural networks to classify semi-classical trajectories \cite{liu_deep_2020}, deep learning to predict in spectra of high-harmonic generation \cite{lytova_deep_2022}, and optimization of ``quantum pathways'' in enhanced ionization of diatomic molecules \cite{chomet_controlling_2022}. In terms of parameter estimation, in a recent theoretical study, CNNs were used to extract internuclear distances and laser intensities, using data generated solving the time-dependent \Schr equation (TDSE) \cite{shvetsov-shilovski_deep_2022, shvetsov-shilovski_transfer_2023}. The power of CNNs to extract useful information from  experimental images has big implications for strong-field physics and attoscience.
Most studies, however, focus on a proof of principle, using only theoretical data. Notable exceptions are, Ref.~\cite{liu_machine_2021}, where CNNs were used to extract molecular structure parameters from experimental laser-induced electron diffraction images, and Ref.~\cite{brunner_deep_2022}, where deep neural networks were applied to streaking traces for parameter extraction and prediction of uncertainties.
Unfortunately, the analytical power of machine learning-assisted imaging is limited if the laser pulse parameters can not be accurately measured.

The characterization of laser pulses in strong-field and attosecond physics has posed a persistent problem. The high intensity of the laser pulse means that a direct measurement (see e.g. \cite{trebino_measuring_1997,kielpinski_benchmarking_2014}) of the intensity leads to significant uncertainties, in the range of $10$--$20\%$ for the strong-field regime \cite{pullen_measurement_2013,kielpinski_benchmarking_2014}. An alternative approach is to use the high sensitivity of the non-linear phenomena in question, to estimate the laser pulse parameters, known as an in situ measurement. Using this approach, laser intensity uncertainties as low as $1\%$ have been reached, by matching experimental results to TDSE theory, under highly controlled experimental conditions for atomic hydrogen \cite{pullen_measurement_2013,kielpinski_benchmarking_2014}. Despite this success, such low uncertainties are not common, and would be more difficult to achieve routinely in standard experimental conditions. Simply fitting photoelectron spectra is less effective and more advanced methods, using all the available information in photoelectron momentum distributions (PMDs), are called for. Recent results, using quantum metrology tools, suggest the uncertainty from in situ measurements could be significantly reduced by exploring quantum interferences present in the PMDs in strong-field ionization \cite{maxwell_quantum_2021}. There is a variety of in situ methods for determining laser parameters that are implemented by hand, whose performances vary across parameter regimes.
The most consistent and powerful method is to use the whole PMD, matching it to that obtained with accurate theoretical methods. As such, ML schemes, and in particular CNNs, are an ideal tool for in situ extraction of laser parameters from experimental data.

The task of using ML for laser pulse characterization has been addressed in the relativistic regime where, a theoretical study used CNNs to predict laser intensities by using proton dynamics \cite{bukharskii_restoration_2023}. In FELs, neural networks have been used to accurately reconstruct pulses, by training a model on a small set of fully diagnosed pulses \cite{sanchez-gonzalez_accurate_2017}. CNNs were also used to characterize the FEL pulses by training on simulated data \cite{ren_temporal_2020}.
CNNs have also been used in all optical measurement schemes, employing interferometric cross-correlation between pulses to characterize one of the pulses \cite{kolesnichenko_neuralnetworkpowered_2023}.
Recently, ML tools have been used for pulse characterization for strong-field ionization \cite{geffert_situ_2022} using purely theoretical data. Here, the autocorrelation function of the ionization yield from two identical pulse was used to extract the pulse duration, spectral width and relative CEP, but the method was insensitive to laser intensity.

In this work, we investigate the power of a CNN as an analysis tool for strong-field physics. We use laser pulse characterization in strong-field ionization as a testing ground, retrieving parameters from PMDs. We train the CNNs on a TDSE model, and test this on large experimental datasets, focusing on retrieving laser intensities, over a larger parameter regime than has previously been considered, for an argon target. The CNNs trained may be used on any experimental data within the parameter range, without special requirements, and the CNN models are available online for testing. Important modifications to theoretical training data, are presented, that ensure the CNN models are insensitive to common experimental imperfections. We also include predictive uncertainty estimation, which goes beyond previous methods to extract uncertainty in strong-field studies.
We produce so-called `explainability' figures that are able to highlight regions of the PMD that contribute the most to a laser intensity prediction, and connect these to the physical interference process that are most sensitive to changes in laser intensity. As such, CNNs represent the easiest way to extract laser parameters in strong-field ionization, making a key step towards producing a general tool for parameters estimation, while also providing more interpretational power. Atomic units are used unless otherwise stated.

\section{Datasets}
\label{sec:datasets}
    \begin{figure*}
        \centering
        \includegraphics[width=\linewidth]{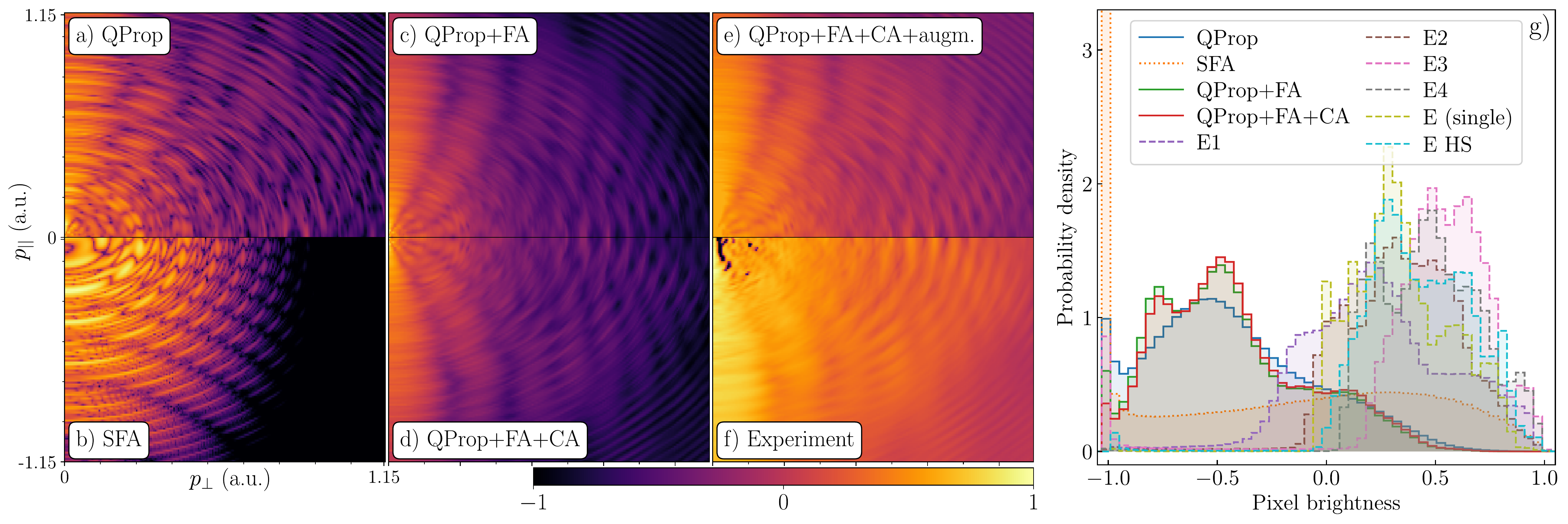}
        \caption{Example PMDs (upper halves in the first row and lower halves in the second row) in the initially preprocessed a) \Qprop, b) SFA ($N=40$ cycles and $U_p=0.3575$) and f) experimental E (single) ($U_p=0.35$) datasets. Color scale corresponds to rescaled and shifted log-probability density and includes the leading 6 orders of magnitude of the original calculated/measured PMD. Panel c) shows the focal-averaged \Qprop\ PMD - minor features are smeared out, panel d) the focal- and CEP-averaged \Qprop\ PMD that does not show qualitative differences from c). Panel e) shows data from d) as seen during training of the model, i.e. at a random detector saturation level $SL=0.4$ (see text for details), contrast $0.8$, brightness $-0.2$. Histograms of pixel brightnesses in panel g) are calculated over a range of pulse ponderomotive potentials $0.15 < U_p < 0.5$. Clearly, direct comparison between theoretical and experimental PMDs is a hard task, as the distributions differ substantially due to experimental limitations.
        }
        \label{fig:ExampleDataSets}
    \end{figure*}
    \subsection{QProp}
        The main workhorse for generating our theoretical data set is the single-active-electron (SAE) TDSE solver \Qprop.
        The latest version of \Qprop~\cite{tulsky_qprop_2019}, implements a fast and accurate method for the calculation of photoelectron momentum distributions (PMDs). \Qprop\ is a velocity gauge 3-dimensional (3D) TDSE solver in the dipole approximation that allows studies within the SAE  approximation using model pseudopotentials\footnote{There is also an implementation of many-electron systems via the solution of the time-dependent Kohn–Sham equations.}. For our SAE model of the Ar atom, we have employed the model potential of Ref.~\cite{tong_empirical_2005}, which has the form
	
	\begin{align}
	    V(r)&=-\frac{Z+f(r)}{r}
	    \intertext{with}
	    f(r)&=a_1 e^{-a_2 r}+a_3 r e^{-a_4 r} + a_5 e^{-a_6 r},
	\end{align}
	where $Z=1$. For argon the coefficients $a_i$ are $a_1=16.039$, $a_2=2.007$, $a_3=-25.543$, $a_4=4.525$, $a_5=0.961$ and $a_6=0.443$~\cite{tong_empirical_2005}, which gives the correct ionization potential of $\ip=0.579$~a.u. In this computation, we considered angular momenta up to $l=55$. Total data set consists of $15712$ PMDs with ponderomotive potentials ranging from $\up=0.0075$ to $\up=0.95$, laser cycles number $N=2$ up to $N=61$. The number of CEP values depends on the pulse length, i.e. for under $N=13$ cycles we cover an interval of length $\pi$ with $10$ values of CEP, and for longer cycles, we have $5$ values of CEP spanning a shifted interval of the same length.
 
    \subsection{Strong Field Approximation}
     An extensive review of the SFA can be found in Ref~\cite{amini_symphony_2019}. Here, we use the transition amplitude for direct ATI from an initial bound state $|\psi_0\rangle$ to a final Volkov state with drift momentum $\pb$ given by \cite{becker_abovethreshold_2002,figueirademorissonfaria_highorder_2002,keldysh_ionization_1965,faisal_multiple_1973,reiss_effect_1980,maxwell_quantum_2021}
    \begin{equation}
    M(\pb)=-i\lim\limits_{t\to\infty}  e^{i S(\pb,t)} \int_{-\infty}^{t}dt' d(\pb,t') e^{i S(\pb,t')},
    \label{eq:SFA-transition-general}
    \end{equation}
    where $d(\pb,t')=\braket{ \pb+\Ab(t')|\rb\cdot\Eb(t')|\Psi_{0}}$ is the dipole prefactor, which in this study we will neglect as we retain only terms correct to exponential accuracy.  The action is given by
    \begin{equation}
        S(\mathbf{p},t)=\ip t +\frac{1}{2}\int^t d t' (\pb+\Ab(t'))^2.
        \label{eq:SFA-action-general}
    \end{equation}
    Here, $\ip$ is the ionization potential of our target.
    We employ the saddle point approximation, seeking the stationary action for the integration variable $t'$, $2\ip+\left(\pb+\Ab(t_s)\right)^2=0$. Now the probability distribution can be computed from \eqref{eq:SFA-transition-general} as
    \begin{align}
        M(\pb)=&
        \sum_s
        c(\pb,t_s,t)	d(\pb,t_s) e^{i S(\pb,t_s)},
        \intertext{where the prefactor $c(\pb,t_s, t)$, derived from the application of the saddle point approximation, includes the $t'$ independent phase from \eqref{eq:SFA-transition-general},  is given by}
        c(\pb,t_s,t)&=-i e^{i S(\pb,t)}\sqrt{\frac{2\pi i}{\partial^2 S(\pb,t_s)/\partial t_s^2}}.
    \end{align}
    In both \Qprop\ and SFA calculations we use the vector potential:
    \begin{equation}
        \Ab(t)=2\sqrt{\up}\sin^2\left( \frac{\omega t}{2 N}\right)\cos(\omega t + \phi),
    \end{equation}
    where $N$ is the number of laser cycles, while $\up$ is the ponderomotive energy or quiver energy of a free electron in the laser field, which is proportional to the peak laser intensity $I_0=2\up c \epsilon_0 \omega^2$, where $\epsilon_0$ and $c$ are the vacuum permittivity and the speed of light, respectively. The angular frequency is given by $\omega$ and the carrier-envelope phase (CEP) is given by $\phi$. We also perform focal (FA) and CEP averaging (CA), to account for variations of the intensity across the focal volume and CEP fluctuations between laser pulses, respectively. CEP-averaged QProp datasets contain $3071$ PMDs. Details are given in the Supplemental material.
    \subsection{Experimental methods}
    \label{sec:experimental_method}

    Our experimental data sets consist of PMDs of argon gas photoionized by intense, linearly polarized, 800~nm laser pulses generated by a 1~kHz commercial Ti:sapphire laser system. These laser pulses are focused in ultra-high vacuum and intersect a pulsed beam of argon gas delivered by an Even-Lavie valve \cite{even_even-lavie_2015}. Photoelectrons are collected in a velocity map imaging spectrometer (VMI), and impact a microchannel plate (MCP) and a fast phosphor detector system. A camera records the intensified phosphor, and on-the-fly peak finding is applied to the live camera feed to extract individual electron impacts.

    Six experimental data sets with $125$ PMDs in total are analyzed here. Four of these sets, labeled E1-E4, were collected using an experimental schema in which the ionizing pulse energy was the variable parameter. The pulse energy is controlled using a motorized rotation mount which manipulates a half-wave plate to rotate the pulse polarization. It then passes through a polarizing beamsplitter cube, which transmits only the component of the laser pulse polarized parallel to the optical table.
    Another data set, “E HS”, was collected for intensities over a larger range and include the highest values of intensity.
    The final set, labeled E (single), is a single-intensity PMD, which is included as it has the highest signal-to-noise ratio.
    Each dataset was collected on the timescale of approximately two days, and contains in total $O(10^9)$ electron counts, distributed between their intensity slices.

    The data sets presented have all been Abel inverted using the standard technique of polar onion peeling \cite{Roberts_2009, werby_dissecting_2021} to extract their cylindrical momentum cross sections. The ponderomotive potential, $U_p$, of the laser fields generating each slice of each data set is computed in a two-step process. First, it is roughly calculated by examining the direct electrons, which form a disk of radius $2U_p$. Then, that rough value is refined by comparing nodes found along the  ``spider-leg" holographic feature to those predicted by the Coulomb quantum orbit strong-field approximation, see Refs. \cite{maxwell_coulombcorrected_2017,maxwell_analytic_2018,maxwell_coulombfree_2018,figueirademorissonfaria_it_2020}, at nearby intensities. This procedure is described in more detail in Ref. \cite{werby_dissecting_2021}. This has an error of approximately $\pm 10\%$.

\section{Deep neural network approach}
	Our task is the following: given an experimental PMD $X$, find a physical parameter $y$, for example, the intensity of the laser pulse, that has been used to produce such PMD. We assume an underlying theoretical model of the strong-field process that allows us to generate PMDs expected at given physical parameters. The richness of the features present in the PMDs makes the task very challenging. Moreover, various imperfections are present in the experimental data, complicating the comparison with theory further, cf. histograms in Fig.~\ref{fig:ExampleDataSets}(g) showing a dramatic quantitative difference between the PMD values in both datasets.

    To address this demanding work, we then use a deep learning approach. By generating a dataset of PMDs labeled by many different physical parameters, we reformulate the problem as a standard supervised regression task. In Section~\ref{sec:cnn}, we describe our choice of deep neural network architecture.

    \subsection{Convolutional Neural Networks and Transfer Learning}
    \label{sec:cnn}
    CNNs \cite{lecun_backpropagation_1989} are designed to work with data incorporating spatial correlations such as pictures. The main building block of CNN is a convolution matrix with trainable parameters that slides over the input image and produces its filtered representation. The composition of many such filters gives a feature map of the image, ready to be used for further processing in the final fully connected part of the network.
    
\begin{figure}
             \centering
             \includegraphics[width=\linewidth]{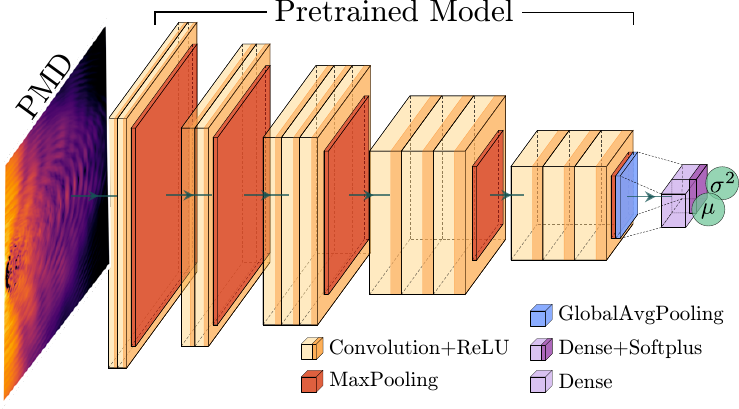}
             \caption{Schematic representation of the Deep Convolutional Neural Network regression problem. For given input $X$ network predicts the value of the parameter $\mu(X)$ and its uncertainty $\sigma(X)$. Adapted from \cite{iqbal_plotneuralnet_2018}.}
             \label{fig:schematic}
         \end{figure}

    Although deep CNNs achieve state-of-the-art in image recognition \cite{krizhevsky_imagenet_2012, russakovsky_imagenet_2014, chen_symbolic_2023}, \changes{sometimes} this requires the use of prohibitively large training datasets and computing power. However, one can train deep models through the transfer learning paradigm \cite{tan_survey_2018, plested_deep_2022}: one takes a pre-trained deep model and fine-tunes it on a smaller dataset. The first few layers of the network extract general features such as edges, so often retraining only the last few layers of the model may be sufficient to achieve good performance on the new dataset. In this work we benchmark four pre-trained architectures called VGG16 \cite{simonyan_very_2015}, Xception \cite{chollet_xception_2017}, EfficientNetB7 \cite{tan_efficientnet_2020}, and EfficientNetV2L \cite{tan_efficientnetv2_2021} that achieved state-of-the-art accuracy in classification tasks on the Imagenet dataset \cite{russakovsky_imagenet_2014}. Models are ordered from least to most sophisticated. We do not describe the substantial innovations introduced in each of them but concentrate on the comparison of their general performance. Because the models were originally designed for classification and not regression, we remove the last classification layer, and replace it with a fully connected layer with a linear (identity) activation function, see Fig.~\ref{fig:schematic}. Models are implemented using Keras library \cite{chollet_keras_2015} and are available in our code repository \url{github.com/tszoldra/attoDNN}.
    
    \subsection{Predictive uncertainty estimation}
    \label{sec:Uncertainty}
    Along with the predicted label, we aim to provide an estimate of the model uncertainty for a given input \cite{gawlikowski_survey_2022, abdar_review_2021}. We slightly modify the model and the loss function \cite{nix_estimating_1994, lakshminarayanan_simple_2017}: instead of predicting a single value of the label $y_{pred}(X)$, we assume the label comes from a normal distribution $p(y_{true}|X) = \mathcal{N}(\mu(X), \sigma(X))$ and the model predicts its parameters $\mu(X), \sigma^2(X)$ for a given input $X$. $\sigma^2(X)$ is the output of an additional fully connected layer with a Softplus$(x) = \ln(\exp(x) + 1)$ activation function, see Fig.~\ref{fig:schematic}. The loss function to minimize is the negative log-likelihood,
    \begin{equation}
    	\text{NLL}\!=\! - \!\left\langle \ln p(y_{true}|X)\right\rangle\!=\!\frac{1}{2}\!\left\langle\! \ln\sigma^2(X) \!+\!\frac{(y_{true}\!-\!\mu(X))^2}{\sigma^2(X)}\!\right\rangle,
     \label{eq:NLL}
    \end{equation}
    where $\langle . \rangle$ denotes the mean over the training dataset. To further \changes{improve} the reliability of \changes{the} predictions and predictive uncertainties, we combine $M$ models trained on random subsets of the training dataset into a deep ensemble \cite{lakshminarayanan_simple_2017, ovadia_can_2019}. Further assuming that the ensemble prediction is a Gaussian, the ensemble mean and variance read
    \begin{equation}
        \mu_*(X) = \frac{1}{M} \sum_{m=1}^M \mu_{m}(X),
    \end{equation}
    \begin{equation}
        \sigma^2_{*}(X) = \frac{1}{M} \sum_{m=1}^M \left( \sigma^2_{m}(X) + \mu^2_{m}(X) \right) -  \mu^2_{*}(X),
        \label{eq:sigma}
    \end{equation}
    where $\mu_m(X),~\sigma_m(X)$ are the mean and variance predicted by the $m$-th model in the ensemble \cite{lakshminarayanan_simple_2017}. This goes beyond studies such as \cite{brunner_deep_2022}, where an ensemble of predictions is used to produce the uncertainty.

        \begin{table*}
\centering
\footnotesize
\begin{tabular}{l|rrrr|rrrr|rrrr|rrrr}
\toprule
 & \multicolumn{4}{c}{VGG16 @ SL=0.0} & \multicolumn{4}{c}{Xception @ SL=-0.5} & \multicolumn{4}{c}{EfficientNetB7 @ SL=0.0} & \multicolumn{4}{c}{EfficientNetV2L @ SL=-0.5} \\
 Dataset& NLL & RMSE & $\langle\sigma_*\rangle$ & MAPE & NLL & RMSE & $\langle\sigma_*\rangle$ & MAPE & NLL & RMSE & $\langle\sigma_*\rangle$ & MAPE & NLL & RMSE & $\langle\sigma_*\rangle$ & MAPE \\\midrule
train & -4.2 & 0.0042 & 0.018 & \textbf{0.77} & -2.0 & 0.11 & 0.12 & 14. & -4.4 & 0.0098 & 0.0097 & 1.3 & -4.1 & 0.015 & 0.012 & 1.7 \\
test & -4.1 & 0.0051 & 0.019 & \textbf{0.87} & -1.9 & 0.12 & 0.12 & 14. & -4.4 & 0.011 & 0.010 & 1.4 & -4.1 & 0.016 & 0.013 & 1.9 \\
E1 & -2.2 & 0.039 & 0.020 & 15. & -1.9 & 0.083 & 0.11 & 33. & 1.7 & 0.038 & 0.010 & \textbf{15}. & -0.99 & 0.040 & 0.016 & 17. \\
E2 & -0.93 & 0.046 & 0.020 & 15. & -2.5 & 0.0099 & 0.083 & \textbf{3.1} & 5.9 & 0.032 & 0.0077 & 12. & -0.70 & 0.034 & 0.012 & 12. \\
E3 & 0.61 & 0.12 & 0.050 & 29. & -2.5 & 0.035 & 0.077 & \textbf{7.2} & -0.89 & 0.037 & 0.021 & 8.5 & -2.3 & 0.043 & 0.029 & 9.8 \\
E4 & -3.4 & 0.021 & 0.015 & 6.2 & -2.2 & 0.061 & 0.089 & 22. & -3.4 & 0.014 & 0.0075 & 4.0 & -4.1 & 0.0091 & 0.014 & \textbf{2.9} \\
E (single) & -3.8 & 0.010 & 0.020 & 2.9 & -2.4 & 0.0072 & 0.087 & 2.0 & -3.1 & 0.015 & 0.0083 & 4.3 & -4.3 & 0.00067 & 0.014 & \textbf{0.19} \\
E HS & 0.70 & 0.11 & 0.041 & 19. & -1.9 & 0.11 & 0.11 & 15. & -2.1 & 0.035 & 0.023 & 4.5 & -3.1 & 0.025 & 0.025 & \textbf{4.5} \\\bottomrule
\end{tabular}

    \caption{Loss metrics for the training dataset QProp+FA+CA. For each CNN architecture, we show only saturation level $SL$ used in the image augmentation for which NLL on dataset E4 is the lowest. On E4, the best accuracy is obtained for EfficientNetV2L which simultaneously gives acceptable errors for other experimental data. Error metrics are similar on test/train subsets of QProp+FA+CA (first two rows) which is a signature of good model generalization. The lowest values of MAPE for each dataset are in bold.}
    \label{tab:models}
\end{table*}

    \subsection{Data preprocessing and augmentation}
    \label{sec:preprocessing}
    In this section, we describe a few technical steps that were necessary to preprocess the PMDs to form a viable image input for the CNNs.
    As a first step, we identified a common range of momenta accessible in all datasets to form a rectangle with $p_\perp \in [0.001 \mathrm{~a.u.}, 1.15 \mathrm{~a.u.}]$, $p_{||}\in [-1.15\mathrm{~a.u.}, 1.15\mathrm{~a.u.}]$, where the discretization is given by the resolution of the experimental dataset E1 $\Delta p_\perp=\Delta p_{||}=0.0049$~a.u. Then, we interpolated the theoretical data on the same 2-dimensional momentum grid.

	The natural scale for features contained in the PMD is logarithmic and the signal has to be transformed accordingly, see eg. \cite{zimmermann_deep_2019} for CNNs applied to diffraction images with similar properties.
    Thus, we take the logarithm of the PMDs, rescale and apply an offset to the pixel values so that in the end they fill the interval $[-1, 1]$. Thus, the pixel value $1$ represents the largest peak probability density in each PMD and $-1$ is the value smaller by a factor of $10^{-6}$. Each pixel is clipped according to $X_{ij} \rightarrow \max (-1, X_{ij})$. Six orders of magnitude are selected based on heuristics that will capture all features in the experimental data, at the same time not misguiding the network by showing extremely precise, low values in the theoretical data. All images are then resized to $224$ by $224$ pixels with $3$ (repeated) color channels to match the standard of the Imagenet dataset expected by the pre-trained models.

	We split the theoretical datasets, see Section \ref{sec:datasets}, into training ($80\%$), validation ($10\%$), and test ($10\%$) subsets. For each model in the ensemble, the training/validation split is different and random, while the test dataset is constructed once by a random selection from the full dataset.

    During the training phase, we perform image augmentation
	\cite{shorten_survey_2019}. Each input image is randomly reflected in the up-down and left-right axes, and its contrast and brightness are randomly set from the interval $(0.1, 1.0)$ and $(-1, 1)$, respectively, using the builtin Keras \cite{chollet_keras_2015} functionalities, see Fig.~\ref{fig:ExampleDataSets}e). The final image is clipped to a fixed range $[-1, 1]$. While the testing data does not include reflected images, during training we apply reflections to help the network learn the same  ``shapes" in four different settings with the aim of reducing overfitting. On the other hand, the testing data has inherently varying levels of contrast and background signal (``brightness") and we deliberately make the network insensitive to them.

    In our efforts to make the models useful for experimentalists, we encountered an obstacle: while the models performed well on theoretical data (see next Section \ref{sec:results}), they failed for experiments. We fixed it by adding a single extra step in the augmentation pipeline, motivated by the histogram in Fig.~\ref{fig:ExampleDataSets}g), that shows significant differences between distributions of pixel values in theory and experiment, especially for the brightest pixels. This suggests there was some uncontrolled detector saturation effect in the experiment, at a level not necessarily fixed between experiments. Thus, prior to all augmentations described earlier, we simulate a random detector saturation level. For each sample PMD, we draw a random  ``saturation" value $x$ from the interval $[SL, 1]$ and transform each pixel according to $X_{ij} \rightarrow \min(X_{i,j}, x) + 1 - x$. The lower bound on the random saturation value $SL$ is a parameter that has to be found by checking the performance of the model on part of the experimental data. By making the saturation level random, we increase the training difficulty, but at the same time make the model insensitive to detector saturation that occurs in a real experiment. \footnote{Since pixel brightness values are limited to the range $[-1,1]$, $SL=-1$ corresponds to a fully random saturation level, whereas $SL=1$ to no detector saturation effect present at all (case of a ``perfect detector").}

\subsection{Training}
	\label{sec:training}
    We train an ensemble of size $M=5$ of pretrained models VGG16, Xception, EfficientNetB7, EfficientNetV2L on four datasets: QProp, QProp+CA, QProp+FA, QProp+CA+FA, for a set of random detector saturation level lower bounds ${SL\in \lbrace -1, -0.5, 0, 0.5, 1 \rbrace}$, yielding $400$ models in total. We choose the Adam optimizer~\cite{kingma_adam_2014} and a batch size of $32$. During the first $50$ training epochs, the base model has fixed weights and only the added, two randomly initialized dense layers each with $1$-dimensional output $\mu(X)$ and $\sigma(X)$ are being updated at a learning rate of $10^{-3}$. This roughly sets up the last layer while not destroying pretrained filters. In the next $150$ iterations the model is fine-tuned: all weights are updated at a learning rate $10^{-4}$, which decreases by a factor of $0.5$ every $50$ iterations. We stop the training if the loss on the evaluation dataset does not decrease for more than $100$ epochs to save on computing time. All models can be trained in parallel.
    \footnote{The training effectiveness may be improved by a recent training scheme \cite{sluijterman_optimal_2023} not applied here: in the first few iterations one should optimize the mean, while keeping the variance fixed.}

    We were unable to train any models if they were initialized with random weights. It demonstrates the power of transfer learning from real-world images to physical experiments. The models are already capable of extracting basic shapes from images and need fine-tuning only.

\section{Results}
	\label{sec:results}
    The quality of all $400$ trained models is measured in terms of the NLL (see \eqref{eq:NLL}), root mean squared error (RMSE) and the mean absolute percentage error (MAPE) achieved on the test datasets. For theoretical data sets, the `true' intensity is known exactly, so RMSE and MAPE give the error on model prediction. For experimental data sets, the `true' intensity carries 10\% uncertainty, so the MAPE only needs to be within this bound.
    The mean predictive uncertainty $\sigma_*(X)$ is the models' prediction of the uncertainty. This can be compared with the RMSE to assess if the models  ``know when they're wrong". All these values are fully tabulated in the Supplemental material. Here we give a general overview of these results and describe a method of model post-selection that allows us to find the best model candidates for experimental data presented in Table~\ref{tab:models}.

    As a standard practice, no augmentation techniques are applied in the testing phase unless explicitly noted. While this can decrease performance of some models on theoretical testing data due to input distribution shifts, we concentrate more on the performance of experimental data which naturally includes imperfections.
    In the “perfect detector” augmentation scenario, $SL=1$, all models are trainable on all theoretical \Qprop\ datasets with a testing MAPE$~<1\%$ (except for the VGG16 model and QProp+CA dataset where MAPE$~=4.9\%$).

    Testing on experimental data, we notice that including focal averaging and CEP-averaging in the training results in a smaller error. This is supported by a visual inspection of the PMDs in Figs.~\ref{fig:ExampleDataSets}a), d), f), which unveils a greater resemblance between experimental and QProp+FA+CA rather than QProp data with more small-sized features. Thus, we restrict our further discussion to the QProp+FA+CA training dataset. Moreover, we observe that the quality of the prediction is always improved if detector saturation effects are included ($SL \leq 0.5$) than if they are not ($SL=1$).

    \begin{figure}[t]
        \centering
            \includegraphics[width=\columnwidth]{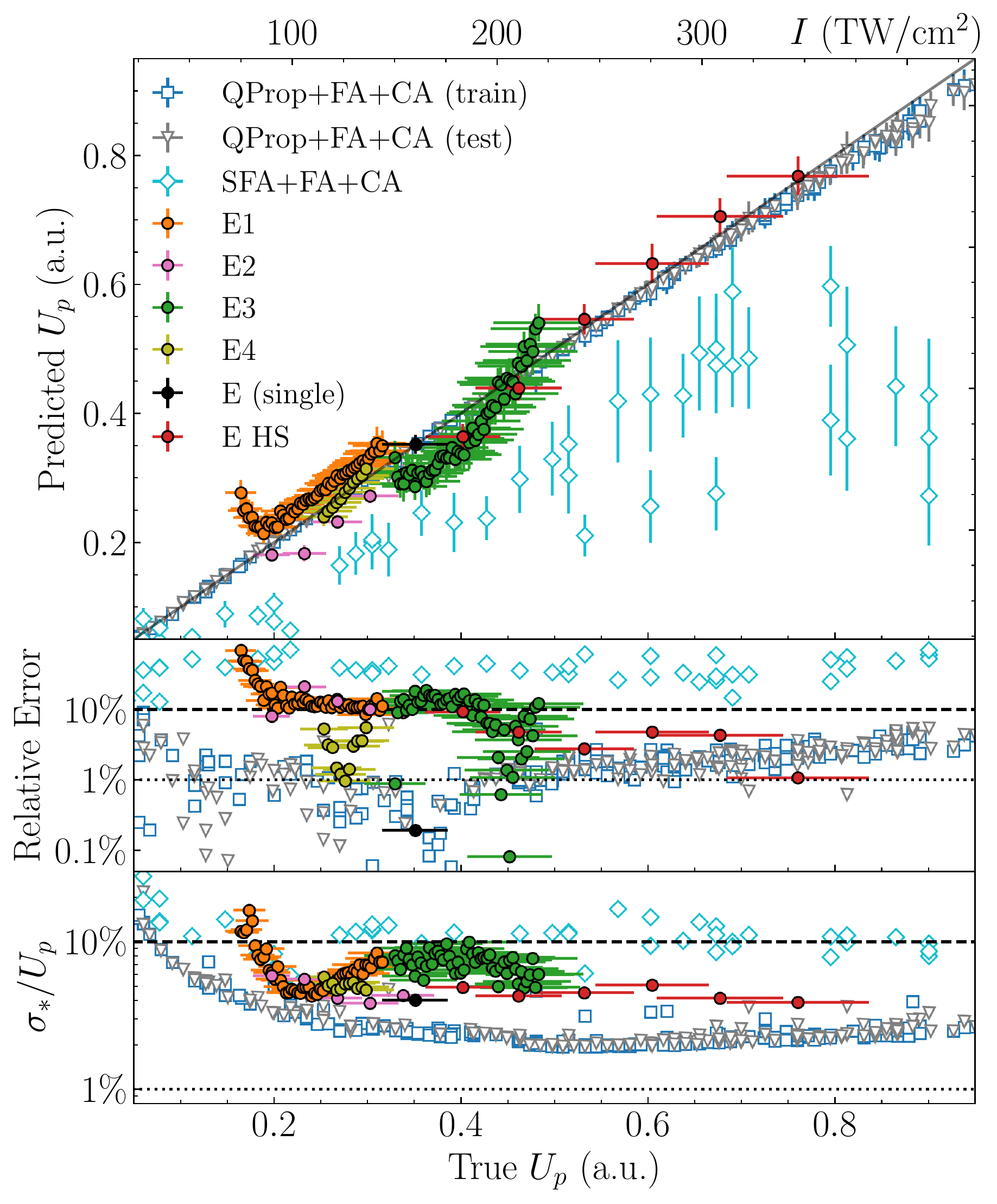}
            \caption{Performance of the EfficientNetV2L model for training dataset QProp+FA+CA, saturation level $SL=-0.5$, for different test datasets. Top plot shows the value of $U_p$ predicted by the model as a function of the true value. Perfect predictions would lie on the black diagonal line. Including focal- and CEP-averaging in the training dataset was necessary to achieve results in agreement with the experimental value, up to the estimated experimental uncertainty level of $10\%$, as shown in the middle plot where most experimental points lie below $10\%$ absolute error line. The bottom plot shows a measure of the model confidence, standard deviation $\sigma_*(U_p)$, as a percentage of the true $U_p$ value.}
            \label{fig:pred_vs_true}
        \end{figure}

   Aiming for the highest-quality models for experiments, we post-select the saturation level based on the best NLL on a single evaluation dataset E4.
   Obtained metrics are presented in Table~\ref{tab:models}. We find that the fittest model is the most sophisticated EfficientNetV2L trained with a saturation level $SL=-0.5$, reaching a MAPE of $2.9\%$ on E4, which is well below the experimental error. The predictive uncertainty is $\sigma_*(X)=0.014$~a.u., corresponding to $6\times 10^{12}$~W/cm$^2$ in typical intensity units. This is close to the reported RMSE, signaling a good calibration of the model confidence. Almost all errors on other datasets for this model also fall within the error bars of the experimental label. In Fig.~\ref{fig:pred_vs_true}, we plot the predicted value of ponderomotive energy $\up$, proportional to the laser intensity, as a function of the true $\up$, for experimental and theoretical (\Qprop\ and SFA) inputs.
   All predictions are presented with uncertainties computed using \eqref{eq:sigma}, given by the vertical error bars in Fig.~\ref{fig:pred_vs_true} (upper panel) and as a percentage (lower panel).

   Predictions on the test/train QProp+FA+CA dataset in Fig.~\ref{fig:pred_vs_true} lie within the uncertainty estimate around the true value up to $U_p\approx 0.5$. For larger $U_p$ the model slightly deviates, finishing with an error of around $7\%$ at $U_p=0.95$. We expect this drop in performance is associated with a limited range of momenta present in the training dataset. The $p=2\sqrt{\up}$ peak, used for labeling PMDs manually, is located at the border of accessible momenta at around $\up\sim 0.66$. Thus, some other, possibly less expressed features in the PMD have to be used by the CNN.

   Before we proceed to experimental data, we quickly cross-check the output of the model on the SFA dataset. Looking at the strong qualitative difference between sample images in Figs.~\ref{fig:ExampleDataSets} a) and b), not to mention the dramatic dissimilarity of the histograms of both datasets in Fig.~\ref{fig:ExampleDataSets}g), it is surprising that the model finds any structure at all in the SFA+FA+CA data, i.e., there is a significant positive correlation coefficient between the true value and the prediction. 
   However, the uncertainty of SFA+FA+CA is strongly underestimated, particularly for the largest values. This is a warning that on the strongly out-of-distribution data\footnote{Out-of-distribution data describes data that is far from the kind of data that was used for training, where the model is likely to give unpredictable results.}, the model may fail to predict the true value and report a relatively high confidence.

   Testing the same model on experimental datasets we notice that errors generally stay equal or lower than the experimental uncertainty of $10\%$ (middle panel). Note that the model was selected based on its performance on evaluation dataset E4, yet its predictions are consistent for other datasets.
   Overall, there is good agreement with the vast majority of intensity predictions carrying an uncertainty below that of attainable through traditional methods, getting as low as 4\% in a number of cases.
   In particular, the high-statistic dataset E HS gives a very good agreement over a wide range of intensities. It was crucial to consider focal averaging here, otherwise, the “true” labels deviated from the predicted value, with an absolute percentage error up to $\sim50\%$ for large intensities.
   Notably, a large relative error is observed for a few points of the dataset E1 at low intensities. We believe this is primarily caused by a low contrast in the input image due to a lower number of electron counts in this setting, see Supplemental material for an example image, since this deviation can be manually removed by increasing the contrast of the input images.

    \section{Explanations}
    \label{sec:explainability}

    \subsection{Methods}

    The high accuracy of the models presented above makes them a readily useful tool for parameter estimation. On the other hand, due to a rather complex flow of information in image regression, the understanding of why a certain output is produced, is lacking, i.e., we deal with a  ``black box”. This hinders progress in the development of new, more accurate models, and, more importantly, does not give any insight into the underlying physical reasoning. These issues are addressed in the following section using so-called explainability techniques that quantify how certain features of the input contribute to the output, see recent review \cite{linardatos_explainable_2020}, or \cite{mohseni_multidisciplinary_2021} for a more general survey.

    The most popular explainability techniques were designed for classifiers, and special care needs to be taken when using them for regression, see \cite{letzgus_toward_2021}. Here we adopt the simple yet powerful strategy of Ref.~\cite{zhang_explainability_2020}, where explanations of deep regression models are obtained directly using methods for classification.

    Three basic approaches to explainable regression have been developed to date and a variety of algorithms can be found in each category \cite{letzgus_toward_2021}. The most straightforward are removal-based explanations \cite{covert_explaining_2021}, measuring the importance of a given subset of input features by hiding it from the model. Because there is an exponential number of subsets to check, usually these methods are limited to at most 15-20 features before the analysis of images becomes infeasible. Another set of methods are gradient-based explanations that rely on the computation of the gradient of the input in a single forward/backward propagation of the signal. They are built on the intuition that if some region of the image is important for the prediction, a small change in this region will noticeably change the output. Finally, propagation-based explanations aim to leverage the neural network structure to produce the feature attribution map. In particular, the layer-wise relevance propagation (LRP) algorithm \cite{bach_pixel-wise_2015} assigns a relevance score $R_i$ to each neuron $i$ based on the activations of the neurons in the next layer. Scores in a single layer are conserved, i.e., sum up to the final prediction. The relevance scores are calculated layer by layer in the backward pass from the output, until the input is reached.

    We apply 10 different explainability algorithms available in the iNNvestigate toolbox \cite{alber_innvestigate_2019}. Our analysis is restricted to the VGG16 model at $SL=0.0$ instead of the  ``best performing” EfficientNetV2L at $SL=-0.5$, presented in Fig.~\ref{fig:pred_vs_true}, due to the large size of the latter, making most methods intractable due to memory requirements, and a  ``swish” activation function which is not compatible with many algorithms. Out of all tested algorithms, for presentation, we post-select the four most relevant ones by scoring them following \cite{samek_evaluating_2017, zhang_explainability_2020}. Each explanation image is divided into $8$ by $8$ regions and perturbed one region at a time, from most to least important, according to the output of a given explanation algorithm. If the regions marked by the algorithm are indeed relevant for predictions, the accuracy of the model drops faster than when the perturbations are applied in random order. In our case, 9 out of 10 tested methods perform better than a random one, and the four presented in Fig.~\ref{fig:explanations} are noticeably better than the rest, see Supplemental Material for further details.

    \subsection{What the model learns.}
    \begin{figure}
        \centering
            \includegraphics[width=1.0\linewidth]{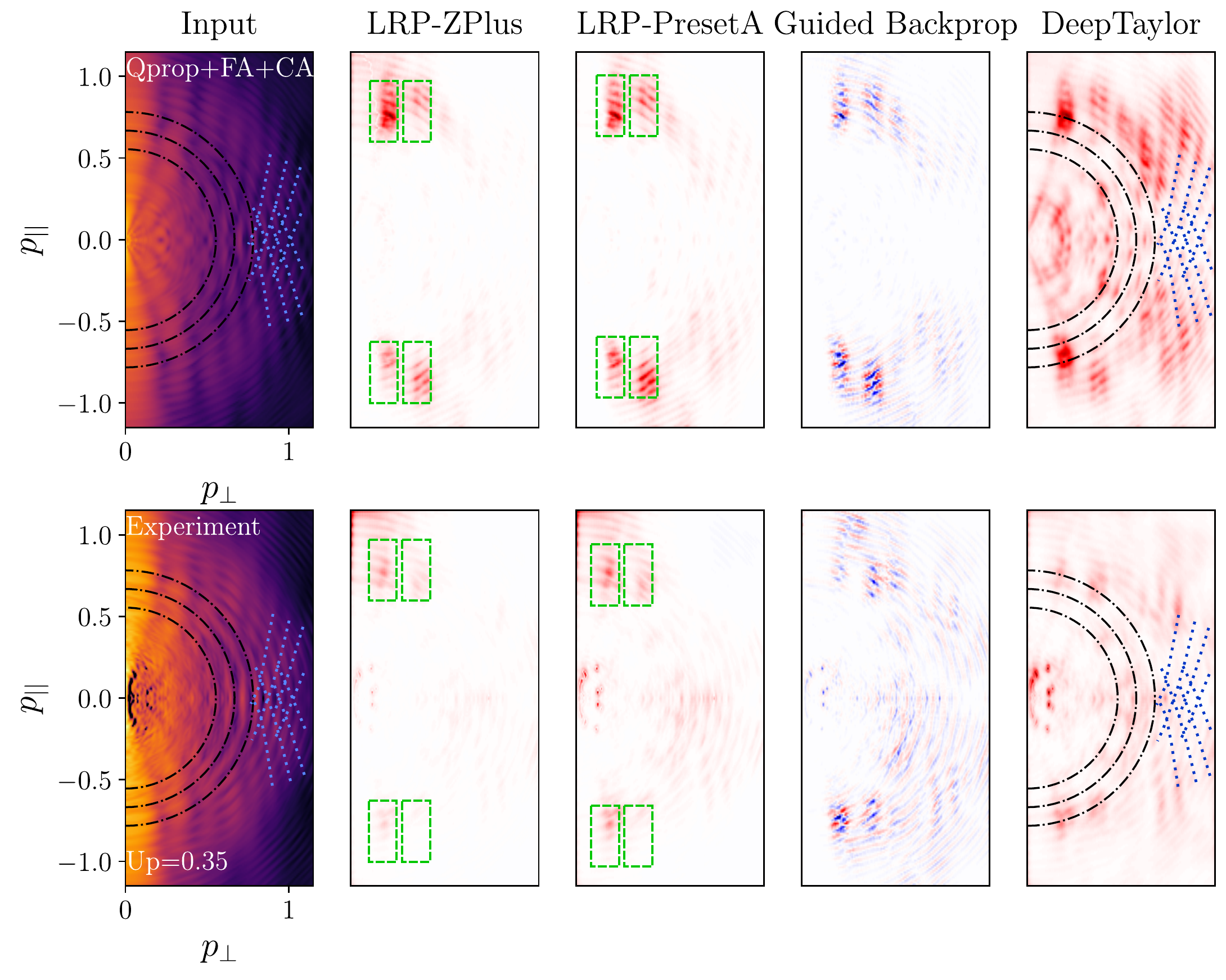}
            \caption{Explanations for the VGG16 model, obtained with 4 most informative methods (upper labels), from left (best) to right (worst). For QProp+FA+CA, true $\up=0.3535$ and predicted $\up=0.3550$, for experiment true $\up=0.351$ and predicted $\up=0.3344$. Red/white/blue colors correspond to positive/neutral/negative attribution in each explainability method.
}
            \label{fig:explanations}
    \end{figure}

    In Fig.~\ref{fig:explanations}, we used four explainability techniques: two variants of the LRP \cite{bach_pixel-wise_2015}, Guided Backpropagation \cite{springenberg_striving_2015}, and DeepTaylor \cite{montavon_explaining_2017} to highlight regions on the PMD that have the most effect on the predicted value. This can also be interpreted as highlighting the features that are the most sensitive to changes in the ponderomotive energy, and thus it is a unique way to extract physical meaning. The highlighted regions can be identified as known interference features that are used in photoelectron holography. The guided backpropagation technique picks out two regions. The first region is at the end of the `legs' of the so-called spider-like structure \cite{huismans_timeresolved_2011,hickstein_direct_2012}, see green dashed rectangles in Fig.~\ref{fig:explanations}. This is formed via the interference between pairs of electronic wavepackets that are forward scattered/deflected off by the residual ion and have a differing degree of interaction with the core.

    The explainability diagrams specifically pick out modulations along the spider legs above the direct boundary ($p=2\sqrt{\up}$). These modulations were discussed in \cite{werby_dissecting_2021}, where their sensitivity to the ponderomotive energy was already exploited for determining the laser intensity.
    Crucially, these modulations have been shown previously \cite{maxwell_coulombcorrected_2017} to be described approximately by circles with their centers determined by $\up$. Thus, explaining the sensitivity to the ponderomotive energy.

    Another region highlighted, with high perpendicular momentum, is the so-called carpet-like \cite{korneev_interference_2012, kang_holographic_2020} or spiral-like \cite{maxwell_spirallike_2020} structure, see blue dotted lines in  Fig.~\ref{fig:explanations}. Around $p_{||}=0$, the interference maxima can be described by $\ip+\up+E=2n\omega$ (with $n\in \mathbb{Z}$), which clearly encodes the ponderomotive energy. Away from $p_{||}=0$, in the DeepTaylor method, we see the strongest contribution. Here, the above equation will not hold exactly, but the fringes will be dependent on the interplay of two rescattered wavepackets, that undergo different rescattering angles, which will be heavily dependent on the laser intensity/ponderomotive energy, as it determines the tunnel exit and initial scattering velocity.

    Across many regions, and particularly in the DeepTaylor method, we can see the above-threshold ionization rings, which are ring-shaped interferences due to nearly identical wavepackets released at an integer number of laser cycles apart, see black dot-dashed circles in Fig.~\ref{fig:explanations}. The maxima may be described by a similar equation to the carpet-like structure $\ip+\up+E=n\omega$, which is clearly sensitive to the ponderomotive energy. In previous work, this has been shown to provide important sensitivity for determining laser intensity \cite{maxwell_quantum_2021}.

    \section{Conclusions}
    We have proposed and tested deep learning models, powerful enough to detect objects in real world images, as a versatile analysis tool for strong-field ionization processes, adopting deep CNNs as our main workhorse. We have found they are capable of extracting physical parameters of interest---the laser peak intensity---and connect the parameter back to a specific feature, such as the interferences observed in the PMDs.
    We have overcome key difficulties using theoretically-trained CNNs with experimental data, which paves the way for CNNs to be used in a variety of settings for strong-field ionization, particularly when characterizing the laser field or target. The CNNs have been tested via pulse characterization, in particular determining the laser field intensity, on a large experimental data set, consistently yielding lower uncertainties than are achievable in traditional methods. For the prediction of uncertainty, we have used a reliable predictive uncertainty approach, that provides additional evaluation of the experimental conditions. We have also verified that other laser field parameters such as the pulse length could easily be extracted.
    
    Deep CNNs can utilize information present in the picture to its full extent, while a human expert would typically be limited to using only a small subset of physical effects that are most sensitive to a change in the parameters. As such, we have developed a novel tool that can be applied to strong-field ionization photoelectron momentum spectra without any special requirements from the experimental data. 
    \changes{We achived this generalizability from our CNN by using state-of-the art pretrained models. Training such models from scratch, to work well with experimental data, was not possible with our data set. Instead using the concept of transfer learning, we found the pretrained networks to work exceptionally well. This approach is easily repeatable as these model are freely availble and it significantly reduces size of the required training data.}
    \changes{The idea of transfer learning may be exploited in future work in order to expand the parameter range of the model, without having to use as much training data. Here, the exceptional capability and large capacity of the CNN models that we use could help make them more generalizable.}
    
    We show-cased the “explainability” capability of CNNs, which highlighted the most relevant features in the PMDs, which could be associated directly with holographic interferences that display considerable sensitivity to changes in the ponderomotive energy. Thus, directly connecting the CNNs predictions to fundamental physical processes.
    \changes{Explainability methods takeaway some of the black box nature of CNNs and can to be used to highligh new physics, providing a powerful way to connect experiment and theory.}

    This study paves the way for further exploitation of CNNs to analyse strong-field ionization data, yielding new physical insights or confirming existing understanding. The recipe we developed, training the neural networks to be insensitive to various types of imperfections through the use of data augmentation techniques, made them ideal candidates for robust parameter extraction. We emphasize that the same procedure can be repeated and used to develop a range of analysis tools, which, for instance, could be highly useful for extracting atomic targets and/or pulse shapes, or further developing photoelectron holographic imaging, where inversion of experimental data is very difficult. Using these techniques, universal extraction of physical parameters is possible from existing and future experimental data, regardless of whether all details of the physical processes at play are fully understood.

	
\acknowledgments
We gratefully acknowledge Poland’s high-performance computing infrastructure PLGrid (HPC Centers: ACK Cyfronet AGH) for providing computer facilities and support within computational grant no. PLG/2022/015830. T.S. is supported by National Science Centre (Poland) under grant 2019/35/B/ST2/00034. This research was
also funded by National Science Centre (Poland) under the OPUS call within the WEAVE programme
2021/43/I/ST3/01142 (J.Z.) A
partial support by the Strategic Programme Excellence
Initiative at Jagiellonian University as well as that within the QuantEra II Programme that has received funding from the European Union's Horizon 2020 research and innovation programme under Grant Agreement No 101017733 DYNAMITE (M.L. and J.Z.). M.F.C. acknowledges financial support from the Guangdong Province Science and Technology Major Project (Future functional materials under extreme conditions, No. 2021B0301030005) and the Guangdong Natural Science Foundation (General Program project No. 2023A1515010871). A.S.M. acknowledges funding support from: The European Union’s Horizon 2020 research and innovation programme under the Marie Sk\l odowska-Curie grant agreement, SSFI No.\ 887153.

N.W. and P.H.B. are supported by the U.S. Department of Energy, Office of Science, Basic Energy Sciences (BES), Chemical Sciences, Geosciences, and Biosciences Division, AMOS Program.

M.L. acknowledges support from: ERC AdG NOQIA; Ministerio de Ciencia y Innovation Agencia Estatal de Investigaciones (PGC2018-097027-B-I00/10.13039/501100011033, CEX2019-000910-S/10.13039/501100011033, Plan National FIDEUA PID2019-106901GB-I00, FPI, QUANTERA MAQS PCI2019-111828-2, QUANTERA DYNAMITE PCI2022-132919, Proyectos de I+D+I “Retos Colaboración” QUSPIN RTC2019-007196-7); MICIIN with funding from European Union NextGenerationEU(PRTR-C17.I1) and by Generalitat de Catalunya; Fundació Cellex; Fundació Mir-Puig; Generalitat de Catalunya (European Social Fund FEDER and CERCA program, AGAUR Grant No. 2021 SGR 01452, QuantumCAT \ U16-011424, co-funded by ERDF Operational Program of Catalonia 2014-2020); Barcelona Supercomputing Center MareNostrum (FI-2022-1-0042); EU Horizon 2020 FET-OPEN OPTOlogic (Grant No 899794); EU Horizon Europe Program (Grant Agreement 101080086 — NeQST), National Science Centre, Poland (Symfonia Grant No. 2016/20/W/ST4/00314); ICFO Internal “QuantumGaudi” project; European Union’s Horizon 2020 research and innovation program under the Marie-Skłodowska-Curie grant agreement No 101029393 (STREDCH) and No 847648 (“La Caixa” Junior Leaders fellowships ID100010434: LCF/BQ/PI19/11690013, LCF/BQ/PI20/11760031, LCF/BQ/PR20/11770012, LCF/BQ/PR21/11840013). Views and opinions expressed in this work are, however, those of the author(s) only and do not necessarily reflect those of the European Union, European Climate, Infrastructure and Environment Executive Agency (CINEA), nor any other granting authority. Neither the European Union nor any granting authority can be held responsible for them.

\providecommand{\noopsort}[1]{} \providecommand{\noopsort}[1]{}

\end{document}